\documentclass[reprint,showkeys,
superscriptaddress,
 amsmath,amssymb,
 aps,prl
]{revtex4-2}

\usepackage{amsmath}
\usepackage{graphicx}
\usepackage{dcolumn}
\usepackage{bm}
\usepackage{graphicx}
\usepackage{dcolumn}
\usepackage{bm}
\usepackage[colorlinks = true, allcolors = blue]{hyperref}
\usepackage{verbatim}
\usepackage{soul}
\usepackage{float}
\usepackage{xcolor}
\usepackage{multirow}
\usepackage{braket}
\usepackage{booktabs}
\usepackage{adjustbox}
\usepackage{appendix}
\usepackage{orcidlink}
\newcommand{\Ry}{\mathrm{Ry}}
\newcommand{\meV}{\,\mathrm{meV}}
\newcommand{\nm}{\,\mathrm{nm}}
\newcommand{\T}{\,\mathrm{T}}

\begin{document}

\preprint{APS/123-QED}

\title{Universal Analytical Description of Magnetoexcitons \\in Two-Dimensional Semiconducting Materials}

\author{Dang-Khoa D. Le\,\orcidlink{0009-0009-8325-0964}}
\affiliation{Computational Physics Key Laboratory K002, Department of Physics, Ho Chi Minh City University of Education, Ho Chi Minh City 72759, Vietnam}

\author{Dai-Nam Le\,\orcidlink{0000-0003-0756-8742}}
\affiliation{Department of Physics and Astronomy and Nebraska Center for Materials and Nanoscience,
\href{https://ror.org/043mer456}{University of Nebraska}, Lincoln, Nebraska 68588, USA}

\author{Ngoc-Loan Phan\,\orcidlink{0000-0003-2292-6157}}
\affiliation{Computational Physics Key Laboratory K002, Department of Physics, Ho Chi Minh City University of Education, Ho Chi Minh City 72759, Vietnam}

\author{Van-Hoang Le\,\orcidlink{0000-0003-4027-0729}}
\email{hoanglv@hcmue.edu.vn}
\affiliation{Computational Physics Key Laboratory K002, Department of Physics, Ho Chi Minh City University of Education, Ho Chi Minh City 72759, Vietnam}

\date{\today}

\begin{abstract}
Matching the Coulomb and logarithmic limits of the Rytova--Keldysh model gives a universal zero-field spectrum for monolayer exciton $s$ states, $E_n\simeq-\mathrm{Ry}^*/[n-1/2+f_n(\xi)]^2$, where $\xi$ is the screening parameter. Analytical radii determine diamagnetic coefficients and the magnetic crossover scale. Zero-field deviations peak near $1\%$ across the full screening range $0\leq\xi\leq100$. For the sampled hBN-encapsulated
monolayers, magnetoexciton energies agree with numerical spectra to sub-meV rms up to $100\T$ and capture the measured orbital shifts. Shared parameters link optical and magneto-optical spectra.
\end{abstract}


\maketitle
\raggedbottom

{\it Introduction --} In 2014, optical spectroscopy revealed Rydberg nonhydrogenic exciton spectra in monolayer WSe$_2$ and WS$_2$ \cite{HePRL2014, chernikov2014}. Optical studies subsequently identified excited exciton states in MoS$_2$, MoSe$_2$, and MoTe$_2$ \cite{Hill2015, Han2018}. High-field magnetoabsorption across these five monolayer materials used diamagnetic shifts to probe exciton radii and follow the crossover from quadratic to nearly linear shifts \cite{Stier2018, NAT2019}. Magneto-photocurrent spectroscopy extended the observed series in WSe$_2$ to $11s$, and pressure-dependent spectroscopy used Rydberg levels to probe changes in dielectric screening \cite{wang2020, Badola2026}. Together, these measurements probe the electron--hole interaction through the zero-field level spacings, their sensitivity to dielectric screening, and the diamagnetic shifts. The Rytova--Keldysh (RK) interaction accounts for the nonhydrogenic spacings through nonlocal dielectric screening. With orbital magnetic confinement included, the same model also describes
the measured diamagnetic shifts \cite{chernikov2014, Stier2016-nano, Stier2016b, Stier2018, NAT2019}.

The RK potential is logarithmic at short electron–hole separations and Coulombic at large separations \cite{Rytova1967, keldysh1979, berkelbach2013}. The logarithmic core depends explicitly on the intrinsic screening length $r_0$. The leading Coulomb tail depends on the environmental dielectric constant $\kappa$ and is independent of $r_0$. At fixed screening, the lowest $s$ states are most affected by the core, whereas higher Rydberg states increasingly sample the tail and approach the two-dimensional hydrogen spectrum at zero field. We therefore focus on the $1s$–$5s$ states in the benchmarks and use both limits to construct an analytical spectrum.

Earlier analytical approaches used a screened hydrogen model, a modified Kratzer potential, or the logarithmic
approximation to the RK interaction \cite{Olsen2016, Molas2019, Hieu2022}. The shifted Rydberg series of Molas {\it et al.} \cite{Molas2019} uses a shift extracted separately for each sample. A regulated perturbation treatment of the full RK interaction gave accurate energies over $0.65 \leq \xi \leq 1.45$, where $\xi$ is the dimensionless screening parameter \cite{Dinh2025}. A related expression was used to extract material parameters \cite{ly2025}. Finite-field studies have used numerical solutions \cite{Stier2018, NAT2019, Nhat2023May} or empirical coefficients determined separately for each material \cite{Nguyen2026}. An open challenge is to obtain a quantitatively accurate analytical description of both zero- and finite-magnetic-field exciton spectra that recovers the Coulomb and logarithmic limits of the RK interaction and uses the same dimensionless coefficients across materials.

Here, we address this challenge with closed-form expressions for the zero-field energies, root-mean-square radii, diamagnetic coefficients, and orbital magnetoexciton energies of exciton $s$ states. Material dependence enters through the reduced mass, intrinsic screening length, and environmental dielectric constant. Matching the two screening limits gives the zero-field energies and radii. The radii fix the diamagnetic coefficients and, with the free inter-Landau-level slopes, set the crossover field scales. A single crossover factor is used across the tested states and materials. We benchmark the zero-field formulas over a broad range of the screening, $0 \leq \xi \leq 100$, and compare the field dependence with numerical RK spectra and polarization-averaged magneto-optical measurements.

{\textit{Model} --} We consider the orbital motion of a neutral exciton with zero center-of-mass pseudomomentum in a perpendicular magnetic field. For $s$ states, the term proportional to relative angular momentum vanishes. With $\rho=r/a_0^*$ and $\varepsilon=E/\Ry^*$, the relative Hamiltonian reads
\begin{equation}\label{eqn1}
  \frac{1}{\Ry^*} {\hat H_s} =-\nabla_{\rho}^{2}+v_\xi(\rho)+\frac{b^2\rho^2}{4},
  \quad b=\frac{\mu_B |B|}{\mu\Ry^*},
\end{equation}
where $\mu$ is the reduced mass in units of the free electron mass $m_e$, $\mu_B=e\hbar/(2m_e)$ is the Bohr magneton, $a_0^*={\kappa}/{\mu}\,a_0$, $\Ry^*=({\mu}/{\kappa^2})\,\Ry$, with the vacuum Bohr radius $a_0$ and Rydberg energy $\Ry$. The Rytova-Keldysh interaction is described by
\begin{gather}\label{eqn2}
  v_\xi(\rho)=-2\int_0^\infty\! \frac{e^{-\rho q}\,dq}{\sqrt{1+\xi^2q^2}}, \quad \xi=\frac{\mu}{\kappa^2}\frac{r_0}{a_0}, 
\end{gather}
where $r_0$ is the intrinsic screening length, and $\kappa$ is the environmental dielectric constant. Equation~\eqref{eqn1} omits valley Zeeman terms. At zero field, $E_n(0)< 0$ is measured from the electron--hole continuum and the binding energy is $-E_n(0)$.

We use regulated perturbation theory (RPT) in a Levi-Civita oscillator basis to guide the asymptotic matching \cite{Dinh2025, giang1993}. The oscillator scale $\omega$ is chosen from the stationary zeroth-order energy for $1s$ and the local minimum of the second-order energy for $2s$--$5s$ in the benchmark calculation. For
hBN-encapsulated WSe$_2$ at $(\mu,r_0,\kappa)=(0.19,4.21\nm,4.34)$, the second-order energies
differ from numerical results by less than $0.60\meV$ and the normalized first-order wave functions have squared overlaps above $99.6\%$. The final interpolations are tested separately against the numerical RK solutions over a wider range of screening.

{\textit{Zero-field energies} --} In the Coulomb limit, $ \xi \ll 1$, the leading screening correction is linear,
\begin{equation}\label{eqn3}
  \varepsilon_n(0)=-\frac{1}{N^2}+\frac{4\xi}{N^3}
                  +{\mathcal{O}}\!\left(\xi^2\ln\frac{1}{\xi}\right),
\end{equation}
where $N=n-\tfrac12$. In the opposite limit, $ \xi \gg 1$, the scaling  $\rho \rightarrow \sqrt{\xi}\,\rho$ separates the constant $-\ln\xi$ from the scaled Hamiltonian. At fixed $n$, this gives
\begin{equation}  \label{eqn4}
  \varepsilon_n(0)=\frac{-\ln\xi+c_n^{\rm log}}{\xi}
                  +{\mathcal{O}}(\xi^{-3/2}),
\end{equation}
where $c_n^{\rm log}$ is the eigenvalue of $-\nabla_{\rho}^2+2\ln(\rho/2)+2\gamma$ and $\gamma$ is Euler's constant. Second-order RPT yields the same asymptotic form with its own estimate of $c_n^{\rm log}$. For $2\leq n\leq50$, semiclassical radial quantization gives constants within $0.82\%$ of the numerical logarithmic eigenvalues. We use this compact semiclassical expression \cite{Gesztesy1978} for the excited states and the numerical value for $1s$,
\begin{equation}\label{eqn5}
  c_n=\begin{cases}
      0.128, & n=1,\\
      2\ln N+2\gamma+\ln(\pi/2), & n\geq2.
  \end{cases}
\end{equation}

Guided by the second-order RPT spectrum, we construct an analytical expression for the zero-field exciton energies that matches the Coulomb and logarithmic screening limits derived above,
\begin{equation}\label{eqn6}
  E_n(0)\simeq-\frac{\Ry^*}{[N+f_n(\xi)]^2},
\end{equation}
where 
\begin{align}
   & f_n(\xi)=\frac{2\xi}{\sqrt{1+4\xi g_n(\xi)}},\label{eqn7}\\
&g_n(\xi)=\ln(1+\xi)-c_n+A_n\ln\!\left(1+\frac{B_n}{\sqrt{\xi}}\right).\label{eqn8}
\end{align}
Here, coefficients $A_n$ and $B_n$ are used to connect the two limits \eqref{eqn3} and \eqref{eqn4}. Using numerical RK energies in the intermediate screening regime for the fitting, we obtain $A_n=2.75-{1.24}/{N}+{0.29}/{N^2}$ and $B_n=11.91+7.56N-0.05N^2$. The expression retains the linear term of Eq.~\eqref{eqn3} and the $-\ln\xi/\xi$ and adopted $c_n/\xi$ terms of Eq.~\eqref{eqn4}. Its large-$\xi$ remainder is ${\mathcal{O}}[(\ln\xi/\xi)^{3/2}]$.

{\textit{Exciton radii and diamagnetic coefficients} --} We use the normalized first-order RPT wave functions to analyze $r_n=\sqrt{\langle r^2\rangle_n}$. We obtain the Coulomb radius and its first screening correction analytically and adopt numerical coefficients for the logarithmic limit. A two-point Pad\'e form fitted to numerical RK radii gives
\begin{equation}\label{eqn9}
  r_n\simeq N^2 \mathcal R_n(\xi)\,a_0^*,
\end{equation}
where
\begin{equation}\label{eqn10}
  \mathcal R_n(\xi)=R_n+
  \frac{L_n(1+5\sqrt{\xi})\xi+C_n^{(r)}\xi^2+D_n^{(r)}\xi^{5/2}}
       {1+5\sqrt{\xi}+25\xi+p_n\xi^{3/2}+q_n\xi^2},
\end{equation}
with the coefficients 
\begin{align}\label{eqn11}
\begin{split}
 R_n = \frac{\sqrt{10+3.5/N^2}}{2},\quad
  L_n = \frac{40+7/N^2}{\sqrt{40N^2+14}},\\
  C_n^{(r)} = \alpha_n p_n+(\beta_n-R_n)q_n,\quad
  D_n^{(r)} = \alpha_n q_n.
\end{split}
\end{align}

Equation~\eqref{eqn10} matches $R_n+L_n\xi$ to first order in weak screening, and the terms adopt $\alpha_n\sqrt{\xi}+\beta_n$ in strong screening. The numerical parameters $\alpha_n$ and $\beta_n$ can be generalized as $\alpha_n = 0.01+{1.83}/{N}+{0.12}/{N^2}$, $\beta_n= 1.52-{0.03}/{N}+{0.17}/{N^2}$. With the RK numerical interpolation, the fitted parameters are   $p_n = 2.89+{27.98}/{N}-{8.19}/{N^2}$ and $q_n = -1.09+{4.54}/{N}+{4.68}/{N^2}$.

The quadratic response follows from Eq.~\eqref{eqn1} with $r_n$ taken from \eqref{eqn9},
\begin{equation}\label{eqn12}
  \sigma_n=\frac{e^2}{8\mu m_e}r_n^2
    \simeq \frac{\kappa^2N^4}{\mu^3}\mathcal R_n^2(\xi)\frac{\mu_B^2}{4\Ry},
  \end{equation}
where $\mu_B^2/(4\Ry)=6.1565\times10^{-5}\,\mu\mathrm{eV/T^2}$. The explicit $N^4$ factor describes the large-$N$ Coulomb limit. At finite screening, $\mathcal R_n$ also depends on $n$, and the mass enters through both the prefactor and $\xi$.

\begin{figure}[ht]
  \centering
  \includegraphics[width=0.36\textwidth]{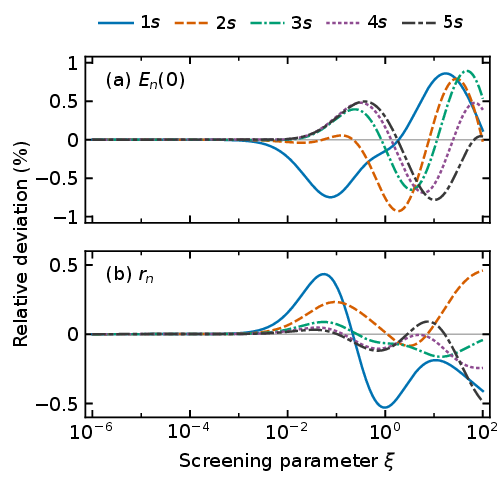}
  \caption{Signed relative deviations of the analytical zero-field energies (a) and rms radii (b) from numerical RK solutions. The test uses 401 screening values from $10^{-6}$ to $100$.}
  \label{Fig1}
\end{figure}

{\textit{Zero-field validation} --} We benchmark the analytical expressions against numerical solutions of the RK Hamiltonian over $10^{-6}\leq\xi\leq100$. This interval extends from the
near-Coulomb regime to strong screening and includes values representative of hBN-encapsulated, substrate-supported, and suspended TMDC monolayers. As shown in Fig.~\ref{Fig1},
the maximum absolute relative deviations across the $1s$--$5s$ states are $0.93\%$ for the zero-field energies and $0.53\%$ for the rms radii. The corresponding maximum
deviation in the diamagnetic coefficients is $1.06\%$, as obtained from $\sigma_n\propto r_n^2$ in Eq.~\eqref{eqn12}.

\begin{figure}[ht]
  \centering
  \includegraphics[width=0.48\textwidth]{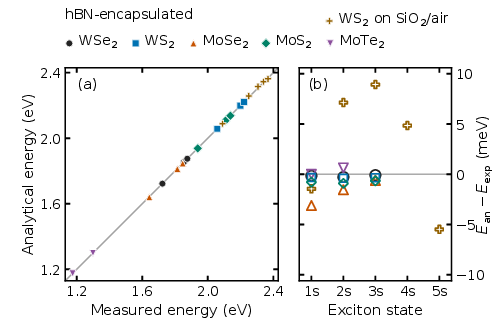}
  \caption{Comparison of analytical and experimental zero-field optical transition energies.  (a) Analytical versus measured energies; the diagonal represents perfect agreement. (b) Energy differences $(E_{\mathrm{an}}-E_{\mathrm{exp}})$ for the corresponding exciton states. Five monolayers are hBN-encapsulated \cite{Stier2018, NAT2019, ly2025}. For WS$_2$ on SiO$_2$/air, we use the effective screening convention of Ref.~\cite{chernikov2014}. The adopted material parameters for analytical calculations are available in Appendix~\ref{appA}.}
  \label{Fig2}
\end{figure}

To compare with optical measurements, we calculate the transition energies as $E_n^{\mathrm{opt}}=E_g+E_n(0)$, using the reported or previously inferred quasiparticle gaps. Figure~\ref{Fig2} compares these analytical results with experimental spectra. The maximum absolute deviation is $3.1 \meV$ for five hBN-encapsulated monolayers and approximately $9 \meV$ for WS$_2$ on SiO$_2$/air, using the effective screening convention of Chernikov {\textit{et al.}}~\cite{chernikov2014}. The larger deviation for supported WS$_2$ may reflect uncertainties in the material parameters and the effective description of its asymmetric dielectric environment. Since some parameters were inferred from the same optical spectra, this comparison primarily tests consistency with the experimental data.

Table~\ref{Tab1} compares the analytical results with values reported from magneto-optical experiments. We use the sample parameters adopted in the original analyses without further
adjustment. For excitons A and B in WS$_2$, the analytical diamagnetic coefficients agree with the measured values within their uncertainties, while the binding energies differ from the inferred values by approximately $5\meV$. For hBN-encapsulated WSe$_2$, the analytical rms radii agree with the values extracted from the magneto-photocurrent spectra within $0.2\%$ for $2s$--$5s$, while the $1s$ radius is approximately $6\%$ smaller.

\begin{table}[ht]
  \caption{Comparison of reported and analytical binding energies (meV), diamagnetic coefficients $(\mu\mathrm{eV/T^2})$, and rms radii (nm). For WS$_2$, the diamagnetic coefficients are measured, while the binding energies and radii are inferred from the diamagnetic response \cite{Stier2016b}. For WSe$_2$, the radii are extracted from numerical fits to magneto-photocurrent spectra \cite{wang2020}. The adopted material parameters for analytical calculations are available in Appendix~\ref{appA}.}
  \label{Tab1}
  \centering
  \footnotesize
  \setlength{\tabcolsep}{3pt}
  \begin{tabular}{@{}lrr@{\qquad}lrr@{}}
    \hline\hline
    \multicolumn{3}{c}{WS$_2$ \cite{Stier2016b}} &
    \multicolumn{3}{c}{WSe$_2$ \cite{wang2020}} \\
    Quantity & Reported & Theory & Radius & Reported & Theory \\
    \hline
    $E_{b,\mathrm{A/B}}$ & 410/470 & \;\,415.0/475.4 & \quad $r_1$ & 1.75 & 1.65 \\
    $\sigma_{1\mathrm A}$ & $0.32\pm 0.02$ & 0.324 & \quad $r_2$ & 6.80 & 6.79 \\
    $\sigma_{1\mathrm B}$ & $0.11\pm0.02$ & 0.109 & \quad $r_3$ & 15.45 & 15.43 \\
    $r_{1\mathrm A}$ & 1.53 & 1.535 & \quad $r_4$ & 27.68 & 27.64 \\
    $r_{1\mathrm B}$ & 1.16 & 1.155 & \quad $r_5$ & 43.51 & 43.44 \\
    \hline\hline
  \end{tabular}
\end{table}

\begin{figure}[ht]
  \centering
  \includegraphics[width=0.47\textwidth]{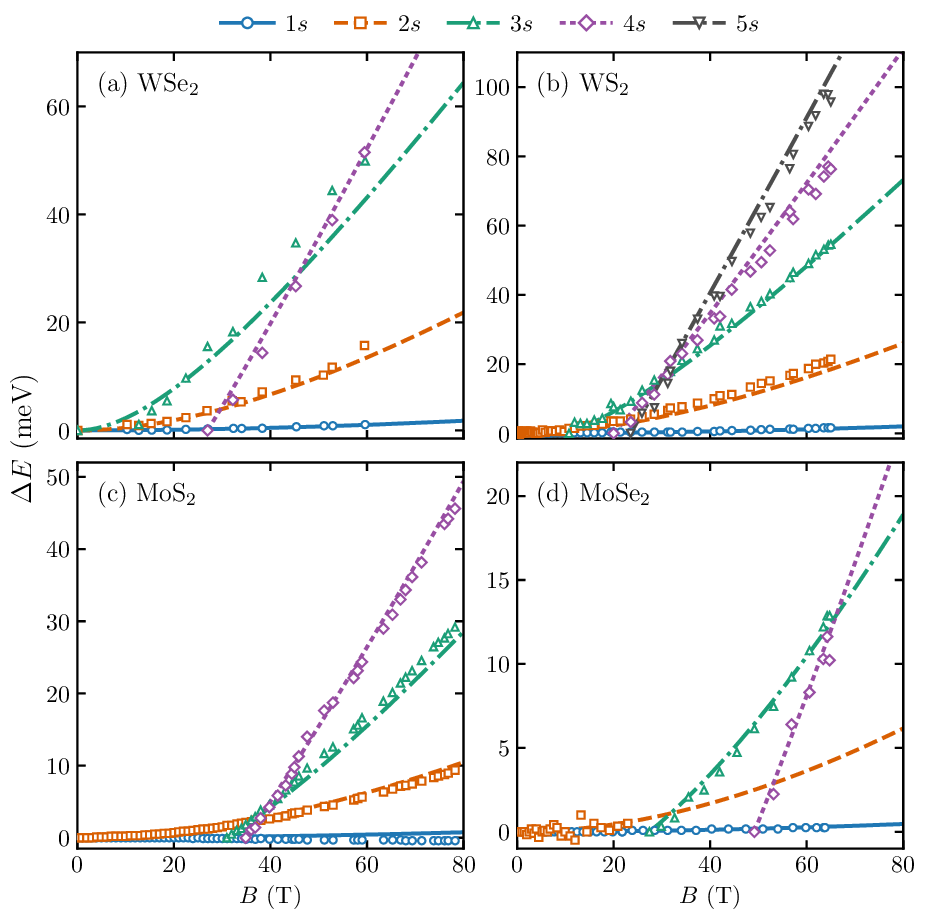}
  \caption{Reference-subtracted exciton shifts $\Delta\bar E_n(B;B_{\rm ref})$ in hBN-encapsulated monolayers. Open symbols denote measurements and lines denote the
  finite-field interpolation. The data are from Stier \textit{et al.}~\cite{Stier2018} for (a) WSe$_2$ and Goryca \textit{et al.}~\cite{NAT2019} for (b) WS$_2$, (c) MoS$_2$, and (d) MoSe$_2$. Each series is referenced to its lowest measured field.}
  \label{Fig3}
\end{figure}

{\textit{Finite-field orbital energies} --} For $B\geq 0$, the weak-field expansion is $E_n(B)=E_n(0)+\sigma_n B^2+\mathcal{O}(B^4)$. At large $B$, the leading term is the $n$-th radial oscillator energy $S_n B$, where $S_n=(2n-1)\mu_B/\mu$. These two limits motivate the Pad\'e interpolation
\begin{equation} \label{eqn13}
  E_n(B)\simeq E_n(0)+
  \frac{\sigma_n B^2}{1+g_{\rm P}\sigma_n B/S_n},
  \quad g_{\rm P}=1.15 .
\end{equation}
Equations~\eqref{eqn6} and \eqref{eqn12} provide the zero-field inputs $E_n(0)$ and $\sigma_n$. The interpolation recovers the weak-field shift $\sigma_n B^2$ and defines a crossover field 
\begin{equation}\label{eqn14}
B_{c,n}=\frac{S_n}{g_{\rm P}\sigma_n}=\frac{(2n-1)\,\mu_B}{g_{\rm P}\mu\sigma_n}, 
\end{equation}
at which the shift is half its quadratic extrapolation. For the WSe$_2$ parameters of Ref.~\cite{Stier2018}, the crossover field decreases from about $820\T$ for $1s$ to $45\T$ for $3s$ and $10\T$ for $5s$, because
$\sigma_n$ grows faster with $n$ than $S_n$. The choice $g_{\rm P}=1$ recovers the exact large-field slope. We calibrate a single value, $g_{\rm P}=1.15$, against numerical RK spectra
over the full screening range considered and for magnetic fields up to $100\T$. We use this value for all five states and all four samples in Fig.~\ref{Fig3}. With this choice, the
interpolation extrapolates to a slope $S_n/g_{\rm P}$ and introduces a cubic term for positive $B$. We therefore use Eq.~\eqref{eqn13} as a finite-field interpolation matched
through order $B^2$.

{\textit{Magneto-optical validation} --} For the WSe$_2$ calibration sample with $\mu=0.20$, $r_0= 4.5 \nm$, and $\kappa= 4.3$, the maximum absolute deviations from numerical RK energies on a $1\T$ grid from $0$ to $100\T$ are $0.28$, $2.05$, $1.37$, $0.61$, and $1.85\meV$ for $1s$--$5s$, respectively. 
For the four hBN-encapsulated samples in Fig.~\ref{Fig3}, which span $0.59\lesssim\xi\lesssim1.33$, the pooled rms deviation from the numerical RK energies on the $1\T$ grid from $0$ to $100\T$ is $0.80\meV$ with $g_{\rm P}=1.15$. The maximum absolute deviation is $2.13\meV$, compared with $30.31\meV$ for $g_{\rm P}=1$. In a broader screening scan over $10^{-2}\leq\xi\leq100$ at fixed $\mu=0.20$ and $\kappa=4.3$, the median absolute relative deviations across the sampled screening values and the states $1s$--$5s$ are around $5.0\%$ at each of $20$, $30$, and $80\T$.

 Figure~\ref{Fig3} compares the finite-field interpolation with measured $ns$ exciton shifts in four hBN-encapsulated monolayers \cite{Stier2018, NAT2019}. We use the mean $\bar E_n=(E_n^{\sigma^+}+E_n^{\sigma^-})/2$ of the two circularly polarized transition energies to remove the linear valley Zeeman contribution. For each state, we take $B_{\rm ref}$ to be the lowest field at which a measured mean energy is available and plot $\Delta\bar E_n(B;B_{\rm ref}) = \bar E_n(B)-\bar E_n(B_{\rm ref})$. The analytical curves use $E_n(B)-E_n(B_{\rm ref})$ from the finite-field interpolation at the same reference field. This subtraction cancels any field-independent energy offset and allows comparison for excited states resolved only at finite fields.

The interpolation agrees closely with the measured shifts for most of the exciton branches shown in Fig.~\ref{Fig3}. For the WSe$_2$ $3s$ branch [Fig.~\ref{Fig3}(a)], the interpolation underestimates the measured shift by about $8.6\meV$ near $53\T$. At the same field, the numerical RK calculation with the same material parameters and reference field underestimates the shift by about $7.4\meV$. Thus, the analytical and numerical RK shifts differ by about $1.2\meV$ at this field.

{\textit{Applications} --} The analytical formulas can guide the choice of dielectric environment and magnetic field for probing exciton states beyond the quadratic regime. For WSe$_2$ at fixed reduced mass and intrinsic screening length, increasing $\kappa$ from $1$ to $4.5$ lowers the crossover field, defined where the shift reaches half its quadratic rule, from hundreds to tens of tesla. The formulas also provide a route to parameter retrieval. At fixed intrinsic screening length and band gap, the zero-field spectrum constrains $\mu/\kappa^2$, while a diamagnetic measurement can distinguish pairs of $\mu$ and $\kappa$ that yield the same spectrum, since $\sigma_n\propto\kappa^{-4}$ at fixed $\mu/\kappa^2$.

{\textit{Conclusion} --} The Coulomb and logarithmic limits of the Rytova–Keldysh model determine the form of a common analytical description of exciton $s$ states. The closed-form energies and radii resolve the screening dependence, while the radii determine the diamagnetic coefficients and supply the state dependence of the magnetic crossover scale. With one shared crossover factor, the resulting orbital energies agree with numerical spectra to $0.80\meV$ rms through 100 T on the grids tested for hBN-encapsulated samples. The same set of formulas thus gives the zero-field spectrum, state sizes, and orbital magnetic response directly from the reduced mass and dielectric screening parameters. These results can guide the choice of dielectric environments and magnetic fields for exciton spectroscopy and help extract material parameters from combined optical and magneto-optical measurements.

{\textit{Acknowledgments} --} This work was supported by the Vietnam National Foundation for Science and Technology Development (NAFOSTED) under Grant No. 103.01-2023.138. The numerical calculations were performed on the high-performance computing cluster at Ho Chi Minh City University of Education, Vietnam.

{\textit{Author contributions} --} V.-H.L. conceptualized the study. D.-K.D.L. and V.-H.L. performed the analytical and numerical calculations, calibrated the analytical formulas, and analyzed the results. D.-N.L. and N.-L.P. contributed to scientific discussions throughout the study and reviewed the manuscript. D.-K.D.L. prepared the initial draft, and V.-H.L. wrote the manuscript. All authors contributed to the final revisions and approved the submitted version.

{\textit{Data availability} --} Data supporting the findings of this study are available from the corresponding author upon reasonable request.

\bibliography{ref}

\appendix
\setcounter{secnumdepth}{1}

\section{Experimental inputs}\label{appA}
\begin{table}[h]
  \caption{Sample parameters for the experimental comparisons. The reduced mass $\mu$ is in units of $m_e$. References identify the sources of the parameters and experimental data.}
  \label{Tab2}
  \centering
  \footnotesize
  \setlength{\tabcolsep}{3pt}
  \begin{tabular}{@{}lcccc c@{}}
    \hline\hline
    Monolayer & $\mu$ & $r_0$ (nm) & $\kappa$ & $E_g$ (eV) & Refs. \\
    \hline
    \multicolumn{6}{@{}l}{\textit{Zero-field optical transitions}, Fig.~\ref{Fig2}} \\
    WSe$_2$ & 0.19 & 4.35 & 4.25 & 1.892 & \cite{Stier2018, ly2025} \\
    WS$_2$ & 0.175 & 3.74 & 4.10 & 2.239 & \cite{NAT2019,ly2025} \\
    MoSe$_2$ & 0.345 & 4.02 & 4.22 & 1.875 & \cite{NAT2019,ly2025} \\
    MoS$_2$ & 0.275 & 3.37 & 4.43 & 2.160 & \cite{NAT2019,ly2025} \\
    MoTe$_2$ & 0.36 & 6.40 & 4.40 & 1.352 & \cite{NAT2019} \\
    WS$_2$ (SiO$_2$/air) & 0.16 & 7.5 & 1.0 & 2.410 & \cite{chernikov2014} \\
    \hline
    \multicolumn{6}{@{}l}{\textit{Binding energies, radii, and diamagnetism}, Tab.~\ref{Tab1}} \\
    WS$_2$ (A exciton) & 0.16 & 5.3 & 1 & -- & \cite{Stier2016b} \\
    WS$_2$ (B exciton) & 0.27 & 5.3 & 1 & -- & \cite{Stier2016b} \\
    WSe$_2$ (radii) & 0.20 & 4.5 & 4.3 & -- & \cite{wang2020} \\
    \hline
    \multicolumn{6}{@{}l}{\textit{Finite-field orbital shifts}, Fig.~\ref{Fig3}} \\
    WSe$_2$ & 0.20 & 4.5 & 4.5 & -- & \cite{Stier2018} \\
    WS$_2$ & 0.175 & 3.4 & 4.35 & -- & \cite{NAT2019} \\
    MoS$_2$ (sample 1) & 0.27 & 3.4 & 4.4 & -- & \cite{NAT2019} \\
    MoSe$_2$ & 0.35 & 3.9 & 4.4 & -- & \cite{NAT2019} \\
    \hline\hline
  \end{tabular}
\end{table}

\end{document}